# Model of work motivation based on happiness: pandemic related study

**Joanna Nieżurawska**[1], **Radosław A. Kycia**[2, 3], **Iveta Ludviga**[4], **and Agnieszka Niemczynowicz**[5, *]

[1]Faculty of Finance and Management, WSB University in Toruń, Poland
[2]Department of Mathematics and Statistics, Masaryk Univeristy, The Czech Republic
[3]Faculty of Computer Science and Telecommunications, Cracow University of Technology, Poland
[4]Riseba University of Applied Sciences, Latvia
[5]Faculty of Mathematics and Computer Science, University of Warmia and Mazury in Olsztyn, Poland
[*]corresponding author: niemaga@matman.uwm.edu.pl

## ABSTRACT

This study aims to enrich the current literature by providing a new approach to motivating Generation Z employees in Poland. Employees need to be motivated in order to be efficient at doing a particular task at the workplace. As young people born between 1995 and 2004 called Generation Z, enter the labour market it, is essential to consider how employees' motivation might be affected. Traditionally identified motivators are known, but many reports indicate that the motivation continues decreasing. This situation causes some perturbations in business and fluctuations of staff. In order to prevent this situation, the employers are looking for new solutions to motivate the employees. A quantitative approach was used to collect new evidence from 200 Polish respondents completing an online survey. The research were conducted before and during pandemic time. We report and analyse the survey results conducted in Poland among representatives of Generation Z, who were employed for at least 6 months. We developed and validated a new approach to motivation using methodologies called Factor Analysis. Based on empirical verification, we found a new tool that connects employee motivation and selected areas of the Hygge concept called *Hygge star model*, which has the same semantics before and during Covid-19 pandemic.

## Introduction

Covid-19 is profoundly affecting almost all aspects of economic and social life globally. Governments have closed borders, banned mass gatherings, and enforced social distancing, generating a new normal for businesses and individual citizens. Measures taken to protect public health have threatened the global economy, necessitating economic stimulus in most countries and reconfiguring the role of business and work value[10]. On December 31, 2019, unexplained pneumonia was first reported to the World Health Organization (WHO). The coronavirus (later named COVID19) was discovered immediately. The threat posed by pandemics, such as Covid-19, has been well known by health organizations and government agencies alike as currently no successful coronavirus vaccine exists. According to the 'Spanish Flu' of 1918, 50 million deaths were the long shadow of the Islands[10]. Just 71 days after the first outbreak, Covid-19 reached 114 countries and resulted in 118,000 cases of and 4,291 deaths[29]. WHO has sounded an alarm and declared Covid-19 a pandemic. Countries were required to take "urgent and aggressive action"[69].

Organizations, regardless of industry and size, strive to build strong and positive relationships with their employees. Changes taking place in the economy, particularly the development of the knowledge-based economy, have forced employers to change their approach to motivating employees. The reason for this is the growing awareness that people are the most valuable resource of the company[70, 72], especially those employees who contribute to the company's progress[12, 31, 40, 63]. Investing in employees manifests itself not only in subsidizing their education, but also in investing in their health, including investing in tools supporting work-life balance, as well as environmental protection and activities related to responsible business. In enterprises, investing in employees is correlated with the implementation of appropriate instruments and concepts of motivating and building a coherent, integrated policy of motivating employees. The policy of motivating employees affects their productivity[4, 7, 19, 62]. Despite the fact that, both in theory and in practice, many systemic solutions for motivating employees have been developed, and companies are outdoing each other in implementing newer instruments and concepts of motivating[2, 18].

In the pandemic time, motivation takes on a new dimension. It turns out that currently, the problem is not just attracting the highly-skilled employees to the organization, but keeping them and motivating them to work. Therefore, the biggest problem for enterprises is maintaining a high level of motivation through the implementation of adequately constructed motivation programs. Previous solutions in the area of motivation have focused on flexibility and individualism. There have been many outstanding scientific papers on the subject, and a number of studies have been conducted both in Poland[46] and in the world[6]. New solutions in this area should go a step further and help shape employees' professional activity, with particular emphasis on the durability of employee attitudes influencing motivation understood in the frame of Hygge concept[8,39]. Taking into account the diversification of employee groups, the concepts should constitute the foundation for creating the tools of motivation. They should be aimed at satisfying the different needs, expectations, and aspirations of employees, building a sense of happiness at work, maintaining a balance between work and private life, as well as material status. In this context, the issue of effective motivating employees is a significant scientific problem, particularly related to young people from Generation Z, who entered the labour market[51]. The problem has only been partially signalled in the literature on the subject and mostly refers to the identifying employees' needs and expectations as part of the motivational activities[21,35,56]. There is a gap in this respect, which has become a premise for undertaking the research presented in this article.

The paper adopts a conventional division into traditional and modern approaches to employee motivation. The available literature of the considered problem, we can find a division into the traditional approach to motivating (e.g.[1,42]) and the modern approach to motivating[17]. Some researchers understand the traditional approach to motivation as remuneration and prestige, often supplemented with meaning, creation, challenge, property, and identity. Furthermore, the modern approach identifies ways of encouraging employees to be more productive and happy simultaneously[17]. Traditional motivation equated with challenging and stimulation, and defined as conscious and deliberate influencing employees to achieve motivator's goals optimally, can be found at work. Concepts based on the traditional approach have undoubtedly made a creative contribution to understanding the phenomenon of work motivation, its sources and mechanisms. The identified needs and expectations, however, did not take into account the diversity and complexity of the contemporary organizational environment, created, among others, by the dynamic development of information technologies, demographic changes (ageing societies), social changes (new behaviour patterns), including generational changes (new needs and expectations, aspirations and priorities regarding life and work). Also innovative entrepreneurs want to increase productivity and well-being by using appreciate motivation tools[24]. Therefore, the traditional approach becomes of little use in solving problems in terms of motivating employees faced by contemporary organizations. The modern approach places a much stronger emphasis on the role of motivation in the labour resources structure. In particular, the age diversity forces a new perspective on the role of work in human life - work that gives a sense of happiness. Taking up the aspects of modern concepts of motivating, taking into account generational diversity, stems mainly from cognitive motives, and has practical justification.

Before Covid-19 organisational success was determined by a high level of employee motivation and managers who properly manage by individuals motivation[17,60] and groups motivation[45]. During the Covid-19 it is noted that managers need to update their strategies and provide a new approach to motivation, especially to the motivation of young people. Expect drastic changes in the upcoming new workforce just entering the labour market. Therefore, the question arises how employers must adapt to the expectations of young people - the first generation which grew up in the era of smartphones and social media, and one which leads a diverse lifestyle during Covid-19.

The objective of this paper is to present a new approach to the employee motivation of young people before and during Covid-19 from theirs point of view. The paper aims is to propose new tools connecting employee motivation and all areas of the Hygge concept, called *Hygge star model* which components have positive influence on employee motivation before and during Covid-19. We discuss different impacts of new approach to motivation before and during Covid-19 of Generation Z. Finally, our research showed that there is a need to introduce a new approach that will increase employees' motivation during the pandemic time. This approach try to answer to expectations of young people from Generation Z[46,47].

The paper is organized as follows. The next section discusses the theoretical models of motivations in the workplace. Further we present empirical part describing methodology and data analysis using factor analysis. Paper ends with conclusions.

## Theories of motivation in the workplace

First theories of motivation were created at the beginning of the 20th century. Since then, numerous research theories of workplace motivation arose throughout the decades. Among them, we can distinguish the needs theories concerned with assessing the needs which the workplace meets or what needs the employee strives to have met[6,11]. Classical theory depiction of human motivation, which we know as the Maslow's hierarchy of needs[42] is based on the assumption that individuals strive to satiate a set of needs and if these needs are met (e.g. through work), then this will lead to greater job satisfaction. In this context we can finds some implications of Malow's theory for managers and some limitations[22,32]. The existence, relatedness, and growth (ERG theory) theory of Adlerfer[3] further developed Maslov's theory. According to Adlerfer the needs can be categorized into three sets: Existence, Relatedness, and Growth. The ERG theory does not assume that the satisfaction of lower order needs is required before pursuing higher order needs[58]. We should note that the results are similar in both theories. The difference is that Maslow was unable to establish empirical evidence and most of the studies were not able to validate his theory[25,36,38,57] either. The next well-known theory of employee motivation was formulated by Frederick Herzberg[28]. The Herzberg's two-factor theory proposes two sets of factors in deciding employees working attitudes and level of performance, named motivation (intrinsic factors) and hygiene factors (extrinsic factors)[58]. Contrary to the previous theories of motivations, this concept focuses on job satisfaction and work results. There are strong correlations between job satisfaction and job results. McClelland[43] identified three motivators: the need for achievement, need for affiliation, and need for power. This approach assumes that people feel all of them at the same time, but with different intensity. The activity is determined depending on the more dominant motivator. It is not essential to achieve other aims. Among many theories of workplace motivation, it is worthy of mentioning these based on the values. Equity theory ([1,64]) states that people prefer or value conditions of fairness and equity in workplace. Valence-instrumentality-expectancy theory (VIE) was first addressed by Vroom[71]. VIE theory was develop by Lawlera and Portera[52]. Researchers proposed that the employees satisfaction depends on achievements. They suggests, that higher achievements determine higher satisfaction. There are two variables, which influence achievements: ability and role perception. It is also observed, that employers and companies are looking for new ways to stimulate their employees towards being more productive and happier at the same time[5]. An example of happiness at work could be the introduction of hygge to the workplace. Jeppe Trolle Linnet[39] described hygge as a social interaction style connected with cultural values. Black and Bodkaer[8] think likewise, seeing hygge as a particular type of social interaction which should be safe and made between people with whom spending time is enjoyable[39,41]. According to Linnet[39], hygge works well in private life as well as in commercial setting[39]. In the latter one, hygge can be applied in most of the human resources management areas, but also in the way the enterprise operates in the business environment. The main areas where hygge can be implemented include: motivating the employees, work organization, but also issues related to the structures within the organization or business activities (cf. Figure 1)[39]. Hygge is a concept of happiness. It can be reflected in theories of needs in the sense that desirable happiness at work can come from meeting various needs. Hygge in the commercial setting is characterized by the presence of greenery, whose aim is to relax and calm down the employees during their work. Another need connected with work is egalitarianism and transparency at the workplace, which translates to equality of all employees and transparency of motivating and remuneration policies. Fair play and lack of aggressive behaviours is yet another need accounted for in the pyramid of employees' needs carried out according to the concept of hygge. The organizational culture of a company which operates according to the concept of hygge is based on mutual trust, teamwork, transparency of actions and decisions, as well as chill, liberty and spontaneity. Hygge also means appropriate working conditions which include a place where a team can eat lunch or have a coffee break. It is important to create a cosy office space, equipped with not only the necessary supplies, but also with abundant greenery, good lighting, and surrounding oneself with personal items - such as a favourite mug on the desk. The need is also visible in creating a cosy and modernly designed office space which is dominated by the colours of nature: earth (brown and green) and the sky (blue). Hygge at the workplace is a style aimed at teamwork, which includes brainstorming, shared problem-solving, conversations, as well as project meetings. According to hygge, the organizational structure of an enterprise should be flat,which fosters transparency and better communication. According to the concept of hygge, motivating the employees is achieved by increasing their engagement thanks to clear and precise goals, proper employee evaluation, and consistent feedback from the employer/manager (cf. Figure 1). Yet another approach to motivating is work-life balance (WLB). It is a specific set of organizational practices, programs and an organizational philosophy that actively helps employees achieve success at work and at home[6]. Work-life balance assumes maintaining harmony between professional and private life. In the world literature on the subject, WLB is understood as a concept, which translates to its definition in this paper[66,68]. Balance between all spheres of functioning is the way to self-realization of the employee, which in turn allows for increased life satisfaction. It is not about the strict separation of the private and professional lives, but rather about their harmonious connection and permeation so that work, family life and passions harmonize with each other. Next theory of motivation is flexible motivation system on the example of the cafeteria. Cafeteria systems of remuneration have developed mainly in the USA, where the wages are relatively high. The remuneration is tailored to the individual needs and expectations of employees, leaving the possibility to choose those elements that best meet their expectations and interests. Cafeteria is often defined as a payroll management tool that confronts the worker with a choice. The cafeteria may include not only the area of benefits, but also deferred income. Bargg defines them as a flexible benefit plan authorized under the Revenue Code. Internal that allows employees to pay for benefits through salary deductions, some of which may be deducted from taxable income[9].

## Methods

### Questionnaire

To examine the motivation system of young people employees in the workplace, an online survey was used. The techniques and technologies used in survey research advanced dramatically during the twentieth century, from systematic sampling methods to improved questionnaire design and compared data analysis. Technology, in particular, has revolutionized how surveys are administered over the last 25 years, with the introduction of the first e-mail survey in the 1980s and the first web-based surveys in the 1990s[20,61]. Although the characteristics of online surveys have been extensively described in the literature[20], online surveys have a number of practical advantages and disadvantages. The online survey was divided into eight sections, but in this study we analyse only one of them (as shown in Appendix **??**), but included two periods of time, before and during Covid-19. The section dealt with a specific piece of information that was required to precisely investigate our research problem.

Only one part of the questionnaire was examined in the current study: this concerning employee motivation systems from the standpoint of their expectations (what are their expectations with regard to the motivation systems). The data were collected in two periods: before Covid-19, it means from October 1st to November 1st, 2020 (the first stage of the research), and during the Covid-19, from May 1st to December 1st, 2021 (the second stage of the research).

This study took into account the following aspects of an online survey: **Q3** (modern systems and concepts of remuneration and motivation, 8 items): The researchers will investigate the significance of a new approach to motivation in this section.

A five-point Likers scale (1 → unimportant, 2 → not so important, 3 → moderately important, 4 → important, 5 → very important) was used for each items. A reliability analysis is conducted from the two constructs used in this study.

From the two constructs used in this study, a reliability analysis is performed. The primary goal of this stage of the research was to investigate the appropriateness of the items and the internal structure of the constructs that the instrument measures.

To test the reliability of the preliminary questionnaire set, reliability on pilot items was performed. The consistency, stability, and dependability of the scores are all factors in an instrument's or questionnaire's reliability[14].

The internal consistency (Cronbach's alpha,[16] were determined to assess reliability. Internal consistency is excellent if the alpha value is greater than 0.9, and acceptable if the alpha value is greater than 0.7. Internal consistency indicates that the survey items tend to cluster together. In other words, a participant who responds positively to one survey item is more likely to respond positively to other survey items.

## Study design

The study was reviewed and approved by the Institutional Review Board and Ethics Commitee of WSB University in Toruń (Poland). Various Internet Platforms used the quatrix[55] to send the proprietary questionnaire. Prizes were distributed among survey participants to encourage potential respondents to complete the questionnaire. The online questionnaire was available in two periods: before Covid-19, that is from October 1st to November 1st, 2020 (the first stage of the research), and during the Covid-19, from May 1st to December 1st, 2021 (the second stage of the research).

## Factor Analysis

In order to establish the validity of the construct, a preliminary study was conducted to determine the unidimensionality of factors with the use of *Exploratory Factor Analysis* (EFA) through a maximum likelihood extraction method[65]. A Equimax rotation has been applied to factorial analysis to identify each variable with a single factor.

Factor analysis is useful in many research areas, e.g. psychology and the social sciences[23]. It helps to get an understanding of the general picture behind the motivation of Generation Z. EFA is a data reduction method applied to a large set of items to identify an underlying factor structure[65]. We applied *Principal Component Analysis* (PCA) extraction with Equimax rotation, which allows for the interpretation of the factor structure. This method allows each item to be loaded highly on one factor and minimize loadings on the remaining factors.

Firstly, we examined the factorability of the part statements related to modern systems and concepts of remuneration and motivation at the workplace. In order to identify the optimum number of factors, we used *the scree test criterion* (Catell's method,[13] or, in the case when this criterion did not work well, *Kaiser criterion*[33,34]. Identification of the factors indicated strong interrelated items of the questionnaire. The Kaiser method assumes the extraction of the factors for which the eigenvalues are greater than one. It suggests that the corresponding factor explains more variance than a single variable[33,34]. The scree test present visual interpretation of the eigenvalues curve of factors. The graph represents the eigenvalues of factors (Y axis) to the corresponding factors (X axis). In this test is that a few major factors account for the most variance. The first, in the graph, we should to find the end of steep "cliff", and count these factors which are above this point. This way we determine the number of extracted factors. Followed the "cliff" by a shallow "scree"[13]. Factorial loads higher than 0.5 were considered as acceptable[65].

The results obtained during the factor analysis were archived and statistically analyzed using Python library FactorAnalyzer[54].

## Participants

The participants of the presented study were employees of Generation Z from Poland. Generation Z consists of young people born between 1995 and 2000.

In the first stage of the study, before Covid-19, sample size was 200, but in the second stage, during Covid-19, 102 participants were qualified. As some research results indicate[30,44,50] not too big sample size (i.e. between 10 and 50) not a large sample size (i.e. between 10 and 50) is needed in order to examine and assess it using factor analysis, as long as communalities are high, the number of expected factors is relatively small, and the model error is low. Based on the mentioned studies, we are certain that the data met the quality criteria necessary to perform the factorial analysis[53]. A factorial analysis of 8 Likert scale questions from this attitude survey was conducted on data gathered from 200 participants from Poland.

The "generational" approach taken by us in the paper discuss the conceptual nature of generations more clearly. We are aware, that some researchers pay attention on how to differentiate generational effects from age or cohort effects[15,48]. However, viewing the employees from the point of generational perspective is of great importance, though not the most crucial factor in providing contextual influences on employee motivation. Conceptual nature of generations should pay some attention on how to differentiate generational effects from age or cohort effects. In order to determine whether a cohort effect is present, we plan a cohort study in the future.

**Ethics statement**

All participants provided informed consent and were informed about the possibility of quitting the survey anytime without consequences. The survey was conducted following the 1964 Declaration of Helsinki with later amendments. Their personal anonymity was preserved (although the relevant data are known to the authors). The survey was distributed by the various Internet Platforms used the quatrix[55] to send the proprietary questionnaire.

The study involves no ethical concerns, and the study materials and design have been approved by the Institutional Review Board (Ethics Committee) WSB University in Toruń, Poland - the approval was issued on 20.20.2019. All methods were performed in accordance with the relevant guidelines and regulations contained in Editorial and publishing policies.

## Results and discussion

**Reliability and validity**

The research problem described in previous sections was investigated using the factor analysis with Equimax rotation. For the part Q3 of the questionnaire containing together 8 items with the grades from 0 to 5, we obtained factors extracted in the Factorial Analysis. The reliability of the Q3 part of the questionnaire is confirmed by computing the Cronbach's alpha. The values are $0,70$ and $0,73$, before and during Covid-19 respectively, which is for basic research. The results show that the constructs used in this research exhibit sufficient internal reliability.

**Factor Analysis results for Q3 before Covid-19**

In this part of the analysis, we examined Generation Z's expectations towards a new approach to motivating companies. According to the scree test criterion and Kaiser method, the Q3 data generated three factors with eigenvalues greater than one, corresponding to $65,3\%$ of the variance explained. Factor 1 (F1) with $31,7\%$ explaining the most variance from other factors, factor 2 (F2) with $16,95\%$ of the variance, and factor 3 (F3) with $16,64\%$ of variance. The Equimax rotation method has been used. In Table 1 a rotated factor matrix is presented.

Variables with higher correlation have been selected for each new factor (see Table 1 and Figure 1):

**F1** (the first factor ) is strongly correlated with the following variables of *the concept of "hygge"*: Q3.4 – Q3.8. It is clearly visible that these variables created the five-arm star (Figure 2). We call it *Hygge star*. Every arm of star corresponds to different area of hygge concept.

**F2** (the second factor) has the variables Q3.2 – *Cafeteria system* with a correlation $0,86$ and Q3.3 – *Flexible remuneration system* with correlation $0,7$. This is the second important factor, with an explained variance of $16,95\%$. Factor 2 can be called *a flexible remuneration system based on the cafeteria systems.*

**F3** (the third factor) is strongly correlated with the variable Q 3.1–*Work-life balance concept* (with $0,83$ correlation), therefore we can define it as *a balance between personal and professional sphere.* This is the third most important factor, with an explained variance of $16,64\%$.

**Factor Analysis results for Q3 during Covid-19**

Now, we present the results concerning Generation Z's expectations towards a new approach how to motivate young people during Covid-19. The scree test criterion and Kaiser method, indicate that the Q3 data generated three factors with eigenvalues greater than one, corresponding to $61,1\%$ of the variance explained. Factor 1 (F1) with $24,1\%$ explaining the most variance from other factors, factor 2 (F2) with $22,3\%$ of the variance, and factor 3 (F3) with $14,7\%$ of variance. We applied the Equimax rotation method as well. The Table 2 presents a rotated factor matrix.

Variables with higher correlation have been selected for each new factor (see Table 2 and Figure 3):

**F1** (the first factor ) is strongly correlated with two variables of *the concept of "hygge"*: Q3.7, Q3.8 and Q3.3 – *Flexible remuneration system*. It is clearly visible that Hygge variables can create the segment, so we call this factoe *Hygge segment* (Figure 4).

**F2** (the second factor) has the three variables of *the concept of "hygge"*, that are Q3.4 – Q3.6. This important factor, with an explained variance of $22,3\%$ can be called *Hygge pyramid*. The figure 5 presents geometrical interpretation of *Hygge pyramid*.

**F3** (the third factor) is strongly correlated with the variable Q 3.1–*Work-life balance concept* (with $0,61$ correlation) and Q3.2 – *Cafeteria system* (with $0,83$ correlation). This factor explained $14,7\%$ of total explained variance.

| Variables | Factors | | |
|---|---|---|---|
| | F1 | F2 | F3 |
| Q3.1: **Work-life balance concept** (keeping balance between your private and professional life) | | | 0,83 |
| Q3.2: **Cafeteria system** (possibility of choosing your own benefits from a list offered by the employer) | | 0,86 | |
| Q3.3: **Flexible remuneration system** (wages are adjusted to the employee's competencies and results) | | 0,7 | |
| Q3.4: **Concept of "hygge"** in the area of designed office space with plants and eco-friendly elements | 0,8 | | |
| Q3.5: **Concept of "hygge"** in the area of flat organizational structure, egalitarianism, and transparency at workplace | 0,80 | | |
| Q3.6: **Concept of "hygge"** in the area fair play and includes not taking aggressive actions on the business market | 0,75 | | |
| Q3.7: **Concept of "hygge"** in the area of organizational culture which includes respect towards one another, teamwork, integration and communication | 0,67 | | |
| Q3.8: **Concept of "hygge"** in the area of role of the manager – leader, who positively motivates the employees, is available to everyone, and is part of the team | 0,58 | | |

**Table 1.** Rotated factor matrix for Q3 before Covid-19.

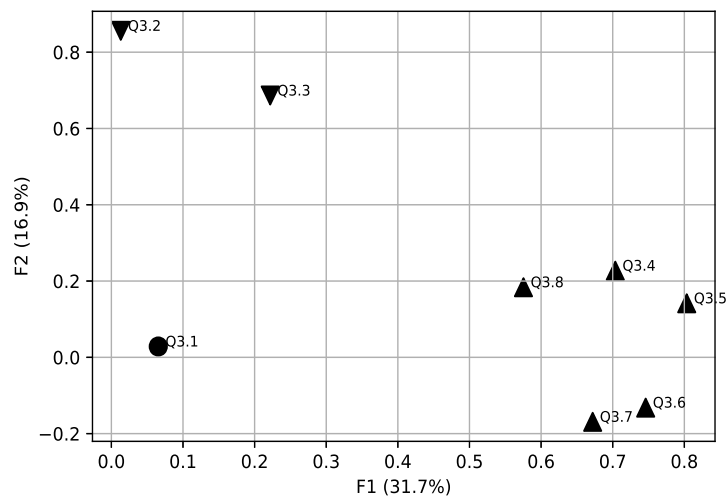

**Figure 1.** 2D factors plot for Q3 data before Covid-19 (correlation matrix FA).

## Limitation

While the authors hope that the study provides some guidance on how to motivate Generation Z, they are aware that it is not without limitations. The first limitation is because the authors focus on relations taking place at one level of analysis, i.e., at the organizational level.

Yet another limitation is looking at employees from a generational perspective. One should be aware that generation is an important, though not the only factor ensuring the effectiveness of motivational strategies. Generations are influenced by extraordinary events that may be radically different across the world. Similar is the case in terms of trends that may take place at different times in individual countries.

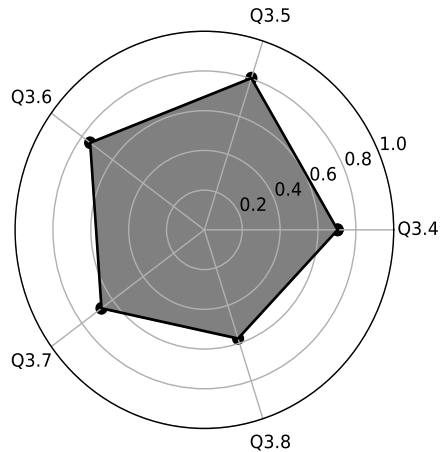

**Figure 2.** Graphical interpretation of the first factor (F1) of Q3 before Covid-19: Hygge Star.

| Variables | Factors | | |
|---|---|---|---|
| | F1 | F2 | F3 |
| Q3.1: **Work-life balance concept** (keeping balance between your private and professional life) | | | 0,61 |
| Q3.2: **Cafeteria system** (possibility of choosing your own benefits from a list offered by the employer) | | | 0,83 |
| Q3.3: **Flexible remuneration system** (wages are adjusted to the employee's competencies and results) | 0,58 | | |
| Q3.4: **Concept of "hygge"** in the area of designed office space with plants and eco-friendly elements | | 0,57 | |
| Q3.5: **Concept of "hygge"** in the area of flat organizational structure, egalitarianism, and transparency at workplace | | 0,78 | |
| Q3.6: **Concept of "hygge"** in the area fair play and includes not taking aggressive actions on the business market | | 0,71 | |
| Q3.7: **Concept of "hygge"** in the area of organizational culture which includes respect towards one another, teamwork, integration and communication | 0,81 | | |
| Q3.8: **Concept of "hygge"** in the area of role of the manager – leader, who positively motivates the employees, is available to everyone, and is part of the team | 0,67 | | |

**Table 2.** Rotated factor matrix for Q3 during Covid-19.

On the other hand, we are aware that the research carried out has a limited territorial scope. The authors plan to conduct further studies with an extended territorial scope (outside Poland) soon.

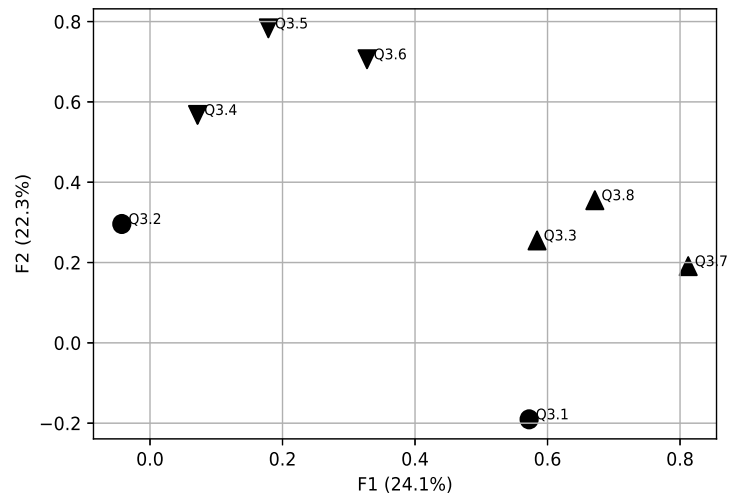

**Figure 3.** 2D factors plot for Q3 data during Covid-19 (correlation matrix FA).

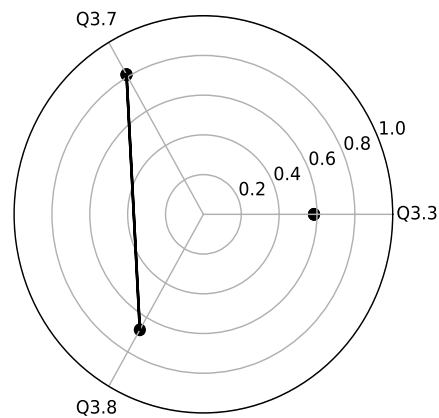

**Figure 4.** Graphical interpretation of the first factor (F1) of Q3 during Covid-19: Hygge segment.

## Implications

At the level of organization it is vital that relevant practitioners such as HR managers note that Generation Z employees:

- require new solutions in the area of motivation;

- are only motivated when there is a good atmosphere in the company, including good relations with the boss and colleagues;

- treat the company as an authority when they receive a lot of support from their boss and are convinced that they have a bright future in this organization;

- are satisfied and the level of their commitment increases when companies implement modern motivational concepts, e.g., hygge;

- should implement the Hygge Star Model, the so-called "Work Happiness Model".

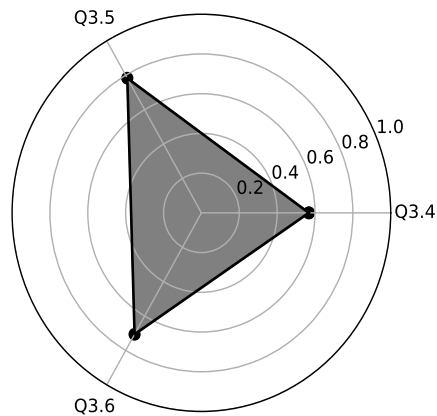

**Figure 5.** Graphical interpretation of the second factor (F2) of Q3 during Covid-19: Hygge pyramid.

At the level of HE institutions, it is important that practitioners such as careers counselors encourage graduates to develop teamwork and brainstorming. Employability programs in HE institutions should focus on helping students find the appropriate company that operates according to the concept of *hygge* based on mutual trust, teamwork, transparency of actions and decisions, as well as chill, liberty, and spontaneity. Practitioners also should notice that hygge in the workplace develops teamwork and brainstorming. Employees look for solutions together. The structure in such an enterprise is flat, which directly improves communication and reduces the distance between colleagues. Our study indicates that, in addition, further development of the hygge concept may be required to develop happiness at work.

## Conclusions

Young workers, in particular, demonstrate high importance regarding the balance between private and professional spheres. For Generation Z, high motivation to work is synonymous with achieving a harmonious and satisfying life in all areas of functioning. They need more time to care for their loved ones and expect to leave the working day in a crisis. The development of work-life balance is influenced by the need for self-fulfillment, which can be achieved thanks to a satisfactory professional situation and a successful private life.

The expectations of Generation Z employees before Covid-19 pandemics in terms of motivation focus on balance between all areas of the Hygge concept. If all areas (star arms) are applicable in an enterprise, there is a high probability that employee motivation will be high or very high. On the other hand, when there is a situation where hygge areas occur to a small extent or not at all, it can be assumed that employees will display a much lower level of employee motivation.

The above assumptions of *the Hygge star model* gave grounds that the Authors perceive the Hygge concept as a *new tool* for measuring the degree of motivation of Generation Z employees at the workplace. In our approach, we assume that the greatest motivation of Generation Z employees occurs when the company has identified all five hygge areas arranged in the structure of a five-pointed, regular star (pentagram). Thus, the ideal situation is when the Hygge star arises as a regular pentagon (the intersection of all diagonals of the regular pentagon). In that case, a balance is maintained between all the hygge areas, as expected by Generation Z employees.

Our new concept of approach to motivation of Generation Z employees is based on empirical studies of young people's expectations, according to which the hygge concept should include all arms (areas) of the Hygge star. If all areas (star arms) are present in the structure of the motivation model of a given enterprise, there is a high probability that employee motivation will be at a high or very high level. On the other hand, in the case of little or no hygge areas, we assume that Generation Z employees will have a much lower level of motivation.

On the other hand, due to changing the work habits to more remote ones because of pandemics, the Hygge star splits into *a segment, a point and a triangle* that vividly indicates the paradigm shift in the concept of motivation. It can understand vaguely as *a simplex*[27], point, segment triangle), used in algebraic topology, that are base elements of the pentagram that appeared before the pandemics (cf. Figure 6).

We proposed the *Hygge pyraimid* as a new approach of motivation that arose during the pandemic time. It consists of three

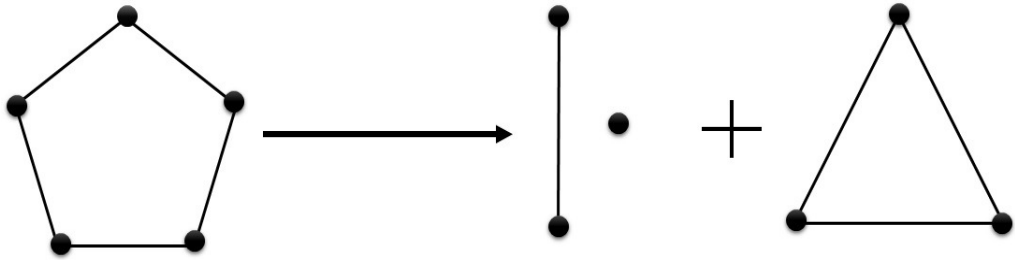

**Figure 6.** The intuitive visualization of the split of the Hygge star (pentagon) during COVID-19 pandemics into simplices (point, segment and triangle). See description in the text.

Hygge components (questions) related to the eco-friendly office design, transparency at the workplace, egalitarianism, and fair play actions. Moreover, for young people, a company with non-aggressive marketing actions is vital to similar concepts.

Another geometric structure we distinguish during pandemic times was named the *Hygge segment* (and a point). The concept is related to a flexible remuneration system, which means that the motivational tools are attached to the competencies of employees and the results of their works. The proposed segment consists only two components (questions) related to Hygge concept: organization culture with teamwork, integration, and communication, and the specific role of manager/leader, who positively motivates the employees and is part of the team. This hidden structure strongly suggests that this approach can ensure high motivation, which can translate to the success of the company. In the future, we plan to extensively investigate such hidden structure interconnections within the Hygge concept.

## Data availability

The datasets used and/or analyzed during the current study are available from the corresponding author on reasonable request. All data generated during this study are included in this article.

## Author contributions

J.N., A.N, conceptualization, methodology and designed the study. J.N., I.L, collected data. A.N., R.A.K., Software development, visualization, data analysis, model checking, supervision. J.N., A.N, R.A.K., writing, funding acquisition. J. N., R.A.K., I. L., and A.N. writing, review. All authors critically reviewed and approved the final version of the manuscript.

## Funding

Funding: This work was supported by the Polish National Agency for Academic Exchange under Grant No. PPI/APM/2019/1/00017/U/00001 project title:*The International Academic Partnership for Generation Z* - 2019-2021.

## Competing interests

The authors declare no competing interests